\documentclass[preprint,12pt]{elsarticle}

\usepackage{amssymb}

\journal{Journal of Computational Physics}

\begin{document}

\begin{frontmatter}




\title{X-Dispersionless Maxwell solver for plasma-based particle acceleration}

\author{Alexander Pukhov}

\address{Institut fuer Theoretische Physik I, Universitaet Duesseldorf, 40225 Germany}

\begin{abstract}
A semi-implicit finite difference time domain (FDTD) numerical Maxwell
solver is developed for full electromagnetic Particle-in-Cell (PIC)
codes for the simulations of plasma-based acceleration. The solver
projects the volumetric Yee lattice into planes transverse to a selected
axis (the particle acceleration direction).
The scheme - by design - removes the numerical dispersion of electromagnetic waves
running parallel the selected axis. The fields locations in the transverse
plane are selected so that the scheme is Lorentz-invariant for relativistic
transformations along the selected axis. The solver results in ``Galilean
shift'' of transverse fields by exactly one cell per time step. This
eases greatly the problem of numerical Cerenkov instability (NCI). The fields
positions build rhombi in plane (RIP) patterns. The RIP scheme uses
a compact local stencil that makes it perfectly suitable for massively
parallel processing via domain decomposition along all three dimensions.
No global/local spectral methods are involved.
\end{abstract}
\maketitle



\end{frontmatter}



\section{Introduction}

Plasma-based particle acceleration is a rapidly developing route towards
future compact accelerators \cite{ALEGRO,Leemans2009,EUpraxia,Joshi_Caldwell}. The
reason is that plasma supports fields orders of magnitude higher than
conventional accelerators \cite{Tajima_Dawson,Esarey_review}. Thus,
particle acceleration can be accomplished on much shorter distances
as compared with the solid-state accelerating strucutures. However,
the plasma is a highly nonlinear medium and requires accurate and
computationally efficient numerical modeling to understand and tune
the acceleration process. The main workhorse for plasma simulations
are Particle-in-Cell codes \cite{OSIRIS,VLPL_CERN,EPOCH,Vay,Smilei,PIConGPU} (a much longerthough still incomplete  list of PIC codes can be found on the web, see e.g. \cite{wikiPIC}).
These provide the most appropriate description of plasma as an ensemble
of particles pushed according to the relativistic equations of motion
using self-consistent electromagnetic fields, which are maintained
on a spatial grid \cite{Birdsall=000026Langdon}.

From a numerical point of view,  plasma-based acceleration represents
a classic multi-scale problem. Here, we have the long scale of acceleration
distance that can range from centimeters \cite{Leemans_8_GeV} to
several meters \cite{SALC,AWAKE}, and the short scale of plasma wavelength
that ranges from a few micrometers to near millimeter scales. In addition,
if the plasma wave is created by a laser pulse, there is additionally
the laser wavelength scale in the sub-micron range. This natural scale
disparity makes the simulations of plasma-based acceleration so computationally
demanding.

Presently, two types of PIC codes are used to simulate the plasma-based
accelerator structures: (i) universal full electromagnetic PIC codes like
\cite{OSIRIS,VLPL_CERN,EPOCH,Vay,Smilei,PIConGPU} , which solve the unabridged
set of Maxwell equations and (ii) quasi-static PIC codes like \cite{WAKE,VLPL_CERN,QUICKPIC,LCODE} (and many others),
which analytically separate the short scale of plasma wavelength and
the long propagation distance scale. The quasi-static PIC codes are
proven to be both accurate and very computationally efficient when
simulating beam-driven plasma-based wake field acceleration (PBWFA).
Unfortunately, the quasi-static aproximation for the Maxwell equations
eliminates any radiation. Thus, the laser pulse driver has to be described
in an envelope approximation \cite{WAKE}. Further, the quasi-static
codes fail at simulating sharp plasma boundaries and self-trapping
of particles from background plasma.

For this reason, we here consider full electromagnetic (EM) PIC codes
which are usually applied for Laser Wake Field Acceleration (LWFA)
in plasmas. The full EM PIC correctly describes the laser evolution
even in highly nonlinear regimes. The full EM PIC codes are computationally
very expensive because they do not separate the different scales.

A significant scale adjustment can be made if one makes a Lorentz
transformation of the system into a reference frame moving in the
direction of acceleration with a relativistic speed. This leads to
the Lorentz contraction of the propagation distance with the relativistic
factor $\gamma=1/\sqrt{1-V^{2}/c^{2}}$, where $V$ is the relative
velocity of the reference frame. Simultaneously, the driver - and
its wavelength - become longer at nearly the same factor. This so-called
``Lorentz-boost'' \cite{Vay_Boost} evens the scale disparity and
potentially gives a large computational speed up at the cost of not properly resolving backward propagating waves.

However, in ``Lorentz-boosted'' PIC simulations, the background
plasma - both electrons and ions - is moving backward at a relativistic
velocity. This moving plasma is a source of free energy that can be
easily transformed into high amplitude noise fields. The major numerical
mechanism for this parasitic conversion is the Cerenkov resonance
\cite{Cerenkov_in_PIC1974}. The problem of most existing FDTD Maxwell
solvers is that they employ the Yee lattice \cite{Yee} (with a few exceptions like FBPIC \cite{FBPIC} and INF\&RNO \cite{INFERNO}): individual
components of the electromagnetic fields are located at staggered positions
in space. The resulting numerical scheme includes a Courant stability
restriction on the time step which leads to numerical dispersion.
This results in electromagnetic waves with phase velocities below
the vacuum speed of light. Thus, the relativistic particles may stay
in resonance with the waves and radiate . This non-physical Cerenkov
radiation plagues the Lorentz-boosted PIC simulations \cite{Vay_Cerenkov_Mitigation}.
Moreover, even normal PIC simulations in the laboratory frame suffer
from the numerical Cerenkov effect \cite{Relativistic_Beam_Cerenkov,Tixon}.
Any high density bunch of relativistic particles - e.g. the accelerated
witness bunch - emits Cerenkov radiation as well. This affects the
bunch energy and emittance \cite{Lehe_Emittance}. 

In principle, the Yee scheme can be modified - or extended - by using
additional neighboring cells with the goal to tune the numerical
dispersion so that the Cerenkov resonance is avoided in the zero order
\cite{Cowan,Cole}. This reduces the Cerenkov instability, but does
not eliminate it. One of the reasons is that the Yee lattice itself
is not Lorentz-invariant. The individual field components are located
all at different positions staggered in space. In the boosted frame,
the fields are Lorentz-transformed and find themselves at the wrong
positions. For example, when the boosted frame moves in the $X-$direction,
the pairs $E_{y},B_{z}$ and $E_{z},B_{y}$ transform one into another.
Yet, they are located at different positions within the Yee lattice
cell. In addition, the aliasing leads to numerical Cerenkov resonances
at wavenumbers from higher Brillouin zones on the numerical grid.

The different positions of the field pairs $E_{y},B_{z}$ and $E_{z},B_{y}$
on the Yee lattice also cause another problem relevant to the high
energy physics. When we want to simulate high current relativistic
beams \cite{NPQED}, this spatial staggering may lead to a beam numerical
self-interaction. A real beam of ultra-relativistic, $\gamma\gg1$,
particles has a small physical self-interaction due to the difference
of these fields with the transverce force $q\left(\mathbf{E_{\perp}+\beta_{||}}\mathbf{e_{||}}\times\mathbf{B}_{\perp}\right)$.
Here $\mathbf{e_{||}}$ is the unit vector in the propagation direction
and $1-\beta_{||}=1-v_{||}/c\approx1/2\gamma^{2}$ is the relative
difference of the particles longitudinal velocity $v_{||}$ from the
speed of light $c$. For $50\,$GeV electrons with $\gamma\approx10^{5}$,
this real difference is as small as $1-\beta_{||}\approx5\cdot 10^{-11}$.
The transverse self-fields ${\bf E}_{\perp}$ and ${\bf B}_{\perp}$
of the ultra-relativistic bunch are also nearly equal with the same
miniscule relative difference. However, the Yee lattice defines these
fields at staggered positions in space and time. These fields must
be interpolated to the same time and to individual particle positions.
This interpolation leads to errors and differences between the transvers
fields acting on the particle. As a consequence, the bunch self-action
due to the numerical errors is many orders of magnitude larger than
the real one. This results not only in the bunch numerical self-focusing/defocusing
and emittance growth, but also in significant numerical bremsstrahlung
and stopping - when these effects are included in the PIC code.  
A similar inaccuracy can occur in the interaction of laser with a co-propagating relativistic beam 
(see the appendix in \cite{LeheLens}).

We conclude, the Yee lattice is not optimal for simulating high energy
applications.

\section{Limitations of pseudo-spectral methods}

Recently, pseudo-spectral methods origninally proposed by Haber et
al. \cite{Haber}, shortly discussed in \cite{Birdsall=000026Langdon}
and used by O. Buneman in his TRISTAN code \cite{Buneman_Tristan}
have seen a remarkable revival \cite{Vay}. The seeming advantage
of the spectral methods is that they are dispersionless and provide
an ``infinite order'' of approximation, even calling the method
after Haber ``a pseudo-spectral analytical time-domain (PSATD) algorithm''
\cite{Vay_convergence}.

Indeed, following Sommerfeld \cite{Sommerfeld} we can write the Maxwell
equations in the Fourier space as 
\begin {eqnarray}
\frac{\partial\hat{\mathbf{F}}}{\partial t} & =ic\mathbf{k}\times\hat{\mathbf{F}}-\hat{\mathbf{J}}
\end {eqnarray}
where $\hat{\mathbf{J}}=FFT[\mathbf{J}]$ is the Fourier image of
the real current while $\hat{\mathbf{F}}=FFT[\mathbf{F}]$ is the
Fourier image of the complex electromagnetic field $\mathbf{F}=\mathbf{E}+i\mathbf{B}$.
It is straightforward to show that the numerical scheme advancing
the fields from the time step $n$ to $n+1$ in the form

\begin {eqnarray}
\hat{\mathbf{F}}^{n+1} & =C_{\mathbf{k}}\hat{\mathbf{F}}^{n}+i\mathbf{S_{k}}\times\hat{\mathbf{F}}^{n}-\tilde{C}_{\mathbf{k}}\hat{\mathbf{J}}^{n+1/2}+i\tilde{\mathbf{S}}_{\mathbf{k}}\times\hat{\mathbf{J}}^{n+1/2}\label{eq:FFT}\\
 & +\left(\tilde{C}_{\mathbf{k}}-1\right)\left(\mathbf{e_{k}}\cdot\hat{\mathbf{J}}^{n+1/2}\right)\mathbf{e_{k}}+\left(1-C_{\mathbf{k}}\right)\left(\mathbf{e_{k}}\cdot\hat{\mathbf{F}}^{n}\right)\mathbf{e_{k}}\nonumber 
\end {eqnarray}
is \textit{dispersionless} in vacuum and provides second order approximation
for the plasma currents. Here, $C_{\mathbf{k}}=\cos(ck\tau)$, $\tilde{C}_{\mathbf{k}}=\cos(ck\tau/2)$,
$\mathbf{S}_{\mathbf{k}}=\mathbf{e_{k}}\sin(ck\tau)$, $\tilde{\mathbf{S}}_{\mathbf{k}}=\mathbf{e_{k}}\sin(ck\tau/2)$,
$\tau$ is the time step, $k=|\mathbf{k}|$, $\mathbf{e_{k}}=\mathbf{k}/k$.

The FFT-based solvers are intrinsically global. This means, they need
information about fields in the full simulation domain to update the
local field at a particular point in space. This contradicts the causality
principle of the special relativity: only fields within $c\tau$ distance
from the space point may cause the local fields to change. The propagator
(\ref{eq:FFT}) explicitly separates the fields and currents into
the propagating transverse fields and non-propagating longitudinal
fields. 

Indeed, the longitudinal part of the field is ${\bf F}_{||}^{n}=\left(\mathbf{e_{k}}\cdot\hat{\mathbf{F}}^{n}\right)\mathbf{e_{k}}$,
and the transverse part is ${\bf F}_{\perp}^{n}={\bf F}^{n}-\left(\mathbf{e_{k}}\cdot\hat{\mathbf{F}}^{n}\right)\mathbf{e_{k}}$, so that ${\bf F}={\bf F_{||}}+{\bf F_{\perp}}$.
The same is valid for the current ${\bf J}={\bf J_{||}}+{\bf J_{\perp}}$.
The Eq. (\ref{eq:FFT}) projected onto the vector ${\bf e_{k}}$ is

\begin{eqnarray}
\hat{F_{||}}^{n+1} & =\hat{F}_{||}^{n}-\hat{J}_{||}^{n+1/2}\label{eq:FFT-1}
\end{eqnarray}
The transverse field components are updated according to

\begin{eqnarray}
\hat{\mathbf{F}}_{\perp}^{n+1} & =C_{\mathbf{k}}\hat{\mathbf{F}}_{\perp}^{n}+i\mathbf{S_{k}}\times\hat{\mathbf{F}}_{\perp}^{n}-\tilde{C}_{\mathbf{k}}\hat{\mathbf{J}}_{\perp}^{n+1/2}+i\tilde{\mathbf{S}}_{\mathbf{k}}\times\hat{\mathbf{J}}_{\perp}^{n+1/2}\label{eq:FFT-2}
\end{eqnarray}
We see that the transverse fields (\ref{eq:FFT-2}) propagate with
the speed of light. The longitudinal component (\ref{eq:FFT-1}) does
not propagate anywhere. Taking divergence of (\ref{eq:FFT-1}) we
arrive at the Poisson equation. 

Computationally, one can use spectral algorithms that are \char`\"{}local\char`\"{}
to one simulation sub-domain \cite{Vay2016,Ultra-High_MAxwell}.
In this case, an ultra-high order finite differences scheme can be
designed. The resulting convolution is then effciently computed within
one sub-domain with the help of a spectral transformation. The resulting
schemes show excellent parallel scalability \cite{Ultra-High_MAxwell}

Although the spectral solver (\ref{eq:FFT}) removes the numerical
dispersion, it does not remove aliasing errors and the numerical Cerenkov
instability in pseudo-spectral codes persists, even when at a lower
rate \cite{Aliasing_PIC,CerenkovFFT}. In an effort to remove the Cerenkov instability
in pseudo-spectral codes, filtering currents of the most unstable
modes is often applied \cite{Cerenko_ Filter,BrendanFilter}. This artificial filtering,
however, may lead to additional unphysical effects in the pseudo-spectral
simulations. 

Another approach is elimination of numerical Cherenkov instability
in flowing-plasma particle-in-cell simulations by using Galilean coordinates
\cite{Lehe_Lagrangian}. This approach removes relative motion between
the numerical grid and the streaming plasma: the grid cells flow together
with the plasma. Unfortunately, this trick works only in one direction
and does not help removing numerical Cerenkov emission of the high
current bunch being accelerated in the opposite direction.

We conclude that pseudo-spectral methods are far from ideal candidates
for PIC simulations and that a better FDTD method is required. In
this work, a new FDTD solver is presented that does not employ spectral
transformations and yet has the unique property of having no numerical
dispersion along one selected spatial axis. 
The positions of the transverse field pairs $\left( E_y, B_z \right)$ $\left( E_z, B_y \right)$  are colocated in the RIP scheme. 
This is Lorentz-invariant and greatly improves the accuracy in calculating the transverse force acting on a relativistic particle moving along the $X-$axis. 

\section{The general X-dispersionless Maxwell solver}

We here develop a FDTD 3D Maxwell solver that has no dispersion for
plane waves propagating in vacuum in one selected direction. In plasma-based
acceleration this is usually the direction of particle acceleration:
the driving laser optical axis. The solver should retain its dispersionless
properties not only in vacuum, but also inside dense plasmas, i.e.
the optimal time step/grid step relation should not be compromised
by the presence of plasma. The solver must not use spectral transformations
and should have a compact local stencil. This is the pre-requisite
for efficient parallelization via domain decomposition. In short,
we develop an efficient Maxwell solver for full three-dimensional problems where one axis is distinguished from the two others (e.g. the laser- or beam-propagation axis). 

We select the $X-$direction for dispersionless propagation. For electromagnetic
waves propagating in $X$, we have the Maxwell equations 

\begin {eqnarray}
\frac{1}{c}\frac{\partial E_{x}}{\partial t} & = & \Gamma_{x} \\
\frac{1}{c}\frac{\partial E_{y}}{\partial t} & = & -\frac{\partial B_{z}}{\partial x}+\Gamma_{y}\\
\frac{1}{c}\frac{\partial E_{z}}{\partial t} & = & \frac{\partial B_{y}}{\partial x}+\Gamma_{z}
\end {eqnarray}

\begin {eqnarray}
\frac{1}{c}\frac{\partial B_{x}}{\partial t} & = & \Phi_{x} \\
\frac{1}{c}\frac{\partial B_{y}}{\partial t} & = & \frac{\partial E_{z}}{\partial x}+\Phi_{y}\\
\frac{1}{c}\frac{\partial B_{z}}{\partial t} & = & -\frac{\partial E_{y}}{\partial x}+\Phi_{z}
\end {eqnarray}

\noindent Here, the vector

\begin {eqnarray}
\Gamma_{x} & = & \frac{\partial B_{z}}{\partial y} - \frac{\partial B_{y}}{\partial z} - J_x\\
\Gamma_{y} & = & \frac{\partial B_{x}}{\partial z} - J_y\\
\Gamma_{z} & = & -\frac{\partial B_{x}}{\partial y} - J_z
\end {eqnarray}

\noindent combines the vacuum diffraction $\mathbf{E}$ and the medium
response (currents) $\mathbf{J}$, while 

\begin {eqnarray}
\Phi_{x} & = & -\frac{\partial E_{y}}{\partial z} + \frac{\partial E_{z}}{\partial y}\\
\Phi_{y} & = & -\frac{\partial E_{z}}{\partial x}\\
\Phi_{z} & = & \frac{\partial E_{y}}{\partial x}
\end {eqnarray}

\noindent is the vacuum diffraction operator for $\mathbf{B}$.

We use a semi-implicit trapezoidal (sometimes called ``implicit midpoint'')
scheme for the discretization of the transverse fields on a 3D grid.
We write here explicitly the $i-$index along the $X-$axis only as
the scheme can be easily generalized for arbitrary transverse geometries
(e.g. Cartesian, or cylindrical, etc.):

\begin {eqnarray}
\frac{E_{y(i+1)}^{n+1}+E_{y(i)}^{n+1}-E_{y(i+1)}^{n}-E_{y(i)}^{n}}{2c\tau} & = &-\frac{-B_{z(i)}^{n+1}+B_{z(i+1)}^{n+1}-B_{z(i)}^{n}+B_{z(i+1)}^{n}}{2h_{x}}\nonumber \\
 & + & \Gamma_{y(i+1/2)}^{n+1/2}\label{eq:Ey} \\
\frac{E_{z(i+1)}^{n+1}+E_{z(i)}^{n+1}-E_{z(i+1)}^{n}-E_{z(i)}^{n}}{2c\tau} & = & \frac{-B_{y(i)}^{n+1}+B_{y(i+1)}^{n+1}-B_{y(i)}^{n}+B_{y(i+1)}^{n}}{2h_{x}}\nonumber \\
 & + & \Gamma_{z(i+1/2)}^{n+1/2} \label{eq:Ez} \\
\frac{E_{x(i)}^{n+1}-E_{x(i)}^{n}}{c\tau} & = & \Gamma_{x(i)}^{n+1/2}\label{eq:Ex}
\end {eqnarray}

\begin {eqnarray}
\frac{B_{y(i)}^{n+1}+B_{y(i+1)}^{n+1}-B_{y(i)}^{n}-B_{y(i+1)}^{n}}{2c\tau} & = & \frac{-E_{z(i)}^{n+1}+E_{z(i+1)}^{n+1}-E_{z(i)}^{n}+E_{z(i+1)}^{n}}{2h_{x}}\nonumber \\
 & + & \Phi_{y(i+1/2)}^{n+1/2}\label{eq:By} \\
\frac{B_{z(i)}^{n+1}+B_{z(i+1)}^{n+1}-B_{z(i)}^{n}-B_{z(i+1)}^{n}}{2c\tau} & = & -\frac{-E_{y(i)}^{n+1}+E_{y(i+1)}^{n+1}-E_{y(i)}^{n}+E_{y(i+1)}^{n}}{2h_{x}} \nonumber \\
 & + & \Phi_{z(i+1/2)}^{n+1/2} \label{eq:Bz} \\
\frac{B_{x(i)}^{n+1}-B_{x(i)}^{n}}{c\tau} & = & \Phi_{x(i)}^{n+1/2}\label{eq:Bx}
\end {eqnarray}
Here, $\tau$ is the time step and $h_{x}$ is the spatial grid step
in the $X-$direction. 

These equations (\ref{eq:Ey})-(\ref{eq:Bx}) build a system of coupled
linear equations relating the updated fields at the time step $n+1$
with already known fields at the time steps $n$ and $n+1/2$. Although
this implicit system of linear equations can generally be solved using
a fast matrix inversion method (the system has a sparse matrix), we
will be interested in the \textbf{special case} $c\tau=h_{x}=\Delta$.
In this particular case, the inversion is straightforward. 

First, we add Eqs. (\ref{eq:Ey})+(\ref{eq:Bz}) and (\ref{eq:Ez})+(\ref{eq:By}).
to obtain transport components

\begin {eqnarray}
T_{y(i)}^{+(n+1)}=E_{y(i)}^{n+1}+B_{z(i)}^{n+1} & = & E_{y(i-1)}^{n}+B_{z(i-1)}^{n}\label{eq:Ey+Bz}\\
 & + & \Delta\left(\Gamma_{y(i-1/2)}^{n+1/2}+\Phi_{z(i-1/2)}^{n+1/2}\right)\nonumber \\
T_{z(i)}^{+(n+1)}=E_{z(i)}^{n+1}+B_{y(i)}^{n+1} & = & E_{z(i+1)}^{n}+B_{y(i+1)}^{n}\label{eq:Ez+By}\\
 & + & \Delta\left(\Gamma_{z(i+1/2)}^{n+1/2}+\Phi_{y(i+1/2)}^{n+1/2}\right)\nonumber 
\end {eqnarray}
or simply 

\begin {eqnarray}
T_{y(i)}^{+(n+1)} & = & T_{y(i-1)}^{+(n)}+\Delta\left(\Gamma_{y(i-1/2)}^{n+1/2}+\Phi_{z(i-1/2)}^{n+1/2}\right)\label{eq:Ty+}\\
T_{z(i)}^{+(n+1)} & = & T_{z(i+1)}^{+(n)}+\Delta\left(\Gamma_{z(i+1/2)}^{n+1/2}+\Phi_{y(i+1/2)}^{n+1/2}\right)\label{eq:Tz+}
\end {eqnarray}
Then, we substract the same Eqs. (\ref{eq:Ey})-(\ref{eq:Bz}) and
(\ref{eq:Ez})-(\ref{eq:By}) to obtain

\begin {eqnarray}
T_{y(i)}^{-(n+1)}=E_{y(i)}^{n+1}-B_{z(i)}^{n+1} & = & E_{y(i+1)}^{n}-B_{z(i+1)}^{n}\label{eq:Ey-Bz}\\
 & + & \Delta\left(\Gamma_{y(i+1/2)}^{n+1/2}-\Phi_{z(i+1/2)}^{n+1/2}\right)\nonumber \\
T_{z(i)}^{-(n+1)}=E_{z(i)}^{n+1}-B_{y(i)}^{n+1} & = & E_{z(i-1)}^{n}-B_{y(i-1)}^{n}\label{eq:Ez-By}\\
 & + & \Delta\left(\Gamma_{z(i-1/2)}^{n+1/2}-\Phi_{y(i-1/2)}^{n+1/2}\right)\nonumber 
\end {eqnarray}
or

\begin {eqnarray}
T_{y(i)}^{-(n+1)} & = & T_{y(i+1)}^{-(n)}+\Delta\left(\Gamma_{y(i+1/2)}^{n+1/2}-\Phi_{z(i+1/2)}^{n+1/2}\right)\label{eq:Ty-}\\
T_{z(i)}^{-(n+1)} & = & T_{z(i-1)}^{-(n)}+\Delta\left(\Gamma_{z(i-1/2)}^{n+1/2}-\Phi_{y(i-1/2)}^{n+1/2}\right)\label{eq:Tz-}
\end {eqnarray}
These are the marching equations. The transport components $T_{y,z}^{+/-}$
must be shifted one cell in the corresponding direction and the diffraction/refraction
terms be correctly added. 

This marching has a form of ``Galiliean field shift'' exactly by
single cell per time step. Thus, instead of shifting the grid following
the relativistic plasma \cite{Lehe_Lagrangian}, the RIP solver shifts
the transverse fields so that the relativistic particle sees the same
fields when it enters the new cell. In one-dimensional geometry, the new algorithm defaults to the well known advective algorithm introduced by Birdsall and Langdon \cite{Birdsall=000026Langdon}. 
In \cite{Cerenkov_in_PIC1974}, and as reported also in \cite{Birdsall=000026Langdon}, on the stability of various electromagnetic PIC schemes, 
it is stated that "the improved stability associated with the advective differencing schemes 
is due not so much to the dispersionless vacuum transport of the fields, per se, 
as to the less conventional methods of determining the mesh current usually employed with advective differencing". 

\begin{figure}
\includegraphics[scale=0.45]{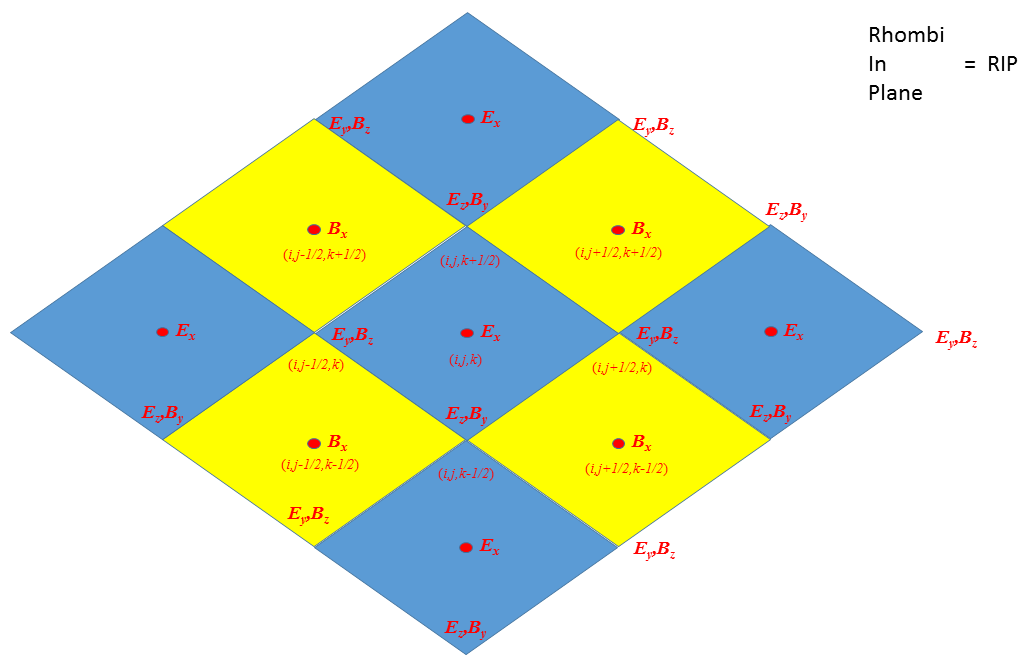}\caption{(color online) The ``rhombi-in-plane'' RIP. grid. \label{fig:RIP}}
\end{figure}

For the fields, we get 

\begin {eqnarray}
E_{y(i)}^{n+1} & =\frac{1}{2}\left(E_{y(i-1)}^{n}+E_{y(i+1)}^{n}\right)-\frac{1}{2}\left(B_{z(i+1)}^{n}-B_{z(i-1)}^{n}\right)\label{eq:Ey+march}\\
 & +\frac{\Delta}{2}\left(\Gamma_{y(i-1/2)}^{n+1/2}+\Phi_{z(i-1/2)}^{n+1/2}+\Gamma_{y(i+1/2)}^{n+1/2}-\Phi_{z(i+1/2)}^{n+1/2}\right)\nonumber \\
E_{z(i)}^{n+1} & =\frac{1}{2}\left(E_{z(i-1)}^{n}+E_{z(i+1)}^{n}\right)+\frac{1}{2}\left(B_{y(i+1)}^{n}-B_{y(i-1)}^{n}\right)\label{eq:Ez+march}\\
 & +\frac{\Delta}{2}\left(\Gamma_{z(i-1/2)}^{n+1/2}-\Phi_{y(i-1/2)}^{n+1/2}+\Gamma_{z(i+1/2)}^{n+1/2}+\Phi_{y(i+1/2)}^{n+1/2}\right)\nonumber 
\end {eqnarray}

\begin {eqnarray}
B_{y(i)}^{n+1} & =\frac{1}{2}\left(B_{y(i-1)}^{n}+B_{y(i+1)}^{n}\right)+\frac{1}{2}\left(E_{z(i+1)}^{n}-E_{z(i-1)}^{n}\right)\label{eq:By+march}\\
 & +\frac{\Delta}{2}\left(-\Gamma_{z(i-1/2)}^{n+1/2}+\Phi_{y(i-1/2)}^{n+1/2}+\Gamma_{z(i+1/2)}^{n+1/2}+\Phi_{y(i+1/2)}^{n+1/2}\right)\nonumber \\
B_{z(i)}^{n+1} & =\frac{1}{2}\left(B_{z(i-1)}^{n}+B_{z(i+1)}^{n}\right)-\frac{1}{2}\left(E_{y(i+1)}^{n}-E_{y(i-1)}^{n}\right)\label{eq:Bz+march}\\
 & +\frac{\Delta}{2}\left(\Gamma_{y(i-1/2)}^{n+1/2}+\Phi_{z(i-1/2)}^{n+1/2}-\Gamma_{z(i+1/2)}^{n+1/2}+\Phi_{y(i+1/2)}^{n+1/2}\right)\nonumber 
\end {eqnarray}
or simply

\begin {eqnarray}
E_{y(i)}^{(n+1)} & =\frac{T_{y(i)}^{+(n+1)}+T_{y(i)}^{-(n+1)}}{2}\label{eq:EyTy}\\
E_{z(i)}^{(n+1)} & =\frac{T_{z(i)}^{+(n+1)}+T_{z(i)}^{-(n+1)}}{2}\label{eq:EzTz}
\end {eqnarray}

\begin {eqnarray}
B_{y(i)}^{(n+1)} & =\frac{T_{y(i)}^{+(n+1)}-T_{y(i)}^{-(n+1)}}{2}\label{eq:EyTy-1}\\
B_{z(i)}^{(n+1)} & =\frac{T_{z(i)}^{+(n+1)}-T_{z(i)}^{-(n+1)}}{2}\label{eq:EzTz-1}
\end {eqnarray}

\section{The three-dimensional RIP Maxwell solver in Cartesian coordinates}

Let us now look at the diffraction/refraction terms. For simplicity,
we use Cartesian coordinates.

We project the Yee lattice onto the $(Y,Z)$ plane. The grid becomes
planar and has the form of Rhombi-in-Plane (RIP), as shown in Fig.~1.
The pairs of transverse fields are now combined at positions according
to the transport properties (\ref{eq:Ey+Bz})-(\ref{eq:Ez+By}). The
pair $\mathbf{E_{y}},\mathbf{B_{z}}$ is located at the rhombi vertices
$(i,j+1/2,k)$. The pair $\mathbf{E_{z}},\mathbf{B_{y}}$ is located
at the rhombi vertices $(i,j,k+1/2)$. The longitudinal field $\mathbf{E_{x}}$
we place at point $(i,j,k)$ which is the center of the full integer
rhombus. The longitudinal field $\mathbf{B_{x}}$ we place at the
center of the half integer rhombus $(i,j+1/2,k+1/2)$. The grid is
shown in Fig.\ref{fig:RIP}.

Then, the diffraction/refraction terms at the half time step will
be:

\begin {eqnarray}
\Gamma_{y(i+1/2,j+1/2,k)}^{n+1/2} & =& \left(\frac{\partial B_{x}}{\partial z}-j_{y}\right)|_{i+1/2,j+1/2,k}^{n+1/2}=-\frac{1}{2}\left(j_{y,(i,j+1/2,k)}^{n+1/2}+j_{y,(i+1,j+1/2,k)}^{n+1/2}\right)+\label{eq:Gy}\\
& + & \frac{B_{x(i,,j+1/2,k+1/2)}^{n+1/2}+B_{x(i+1,,j+1/2,k+1/2)}^{n+1/2}-B_{x(i,j+1/2,k-1/2)}^{n+1/2}-B_{x(i+1,j+1/2,k-1/2)}^{n+1/2}}{2h_{z}}\nonumber \\
\Gamma_{z(i+1/2,j,k+1/2)}^{n+1/2} & = & \left(-\frac{\partial B_{x}}{\partial y}-j_{y}\right)|_{i+1/2,j,k+1/2}^{n+1/2}=-\frac{1}{2}\left(j_{z,(i,j,k+1/2)}^{n+1/2}+j_{z,(i+1,j,k+1/2)}^{n+1/2}\right)-\label{eq:Gz}\\
& - & \frac{B_{x(i,,j+1/2,k+1/2)}^{n+1/2}+B_{x(i+1,j+1/2,k+1/2)}^{n+1/2}-B_{x(i,j-1/2,k+1/2)}^{n+1/2}-B_{x(i+1,j-1/2,k+1/2)}^{n+1/2}}{2h_{y}}\nonumber \\
\Gamma_{x(i,j,k)}^{n+1/2} & =& \left(\frac{\partial B_{z}}{\partial y}-\frac{\partial B_{y}}{\partial z}+j_{x}\right)|_{i,j,k}^{n+1/2}=-j_{x,(i,j,k)}^{n+1/2}\label{eq:Gx}\\
& + & \frac{B_{z(i,,j+1/2,k)}^{n+1/2}-B_{z(i,,j-1/2,k)}^{n+1/2}}{h_{y}}-\frac{B_{y(i,,j,k+1/2)}^{n+1/2}-B_{y(i,,j,k-1/2)}^{n+1/2}}{h_{z}}\nonumber 
\end {eqnarray}
\begin {eqnarray}
\Phi_{y(i+1/2,j,k+1/2)}^{n+1/2} & = &-\frac{E_{x(i,,j,k+1)}^{n+1/2}+E_{x(i+1,,j,k+1)}^{n+1/2}-E_{x(i,j,k)}^{n+1/2}-E_{x(i+1,j,k)}^{n+1/2}}{2h_{z}}\label{eq:Fy}\\
\Phi_{z(i+1/2,j+1/2,k)}^{n+1/2} & = &\frac{E_{x(i,,j+1,k)}^{n+1/2}+E_{x(i+1,,j+1,k)}^{n+1/2}-E_{x(i,j,k)}^{n+1/2}-E_{x(i+1,j,k)}^{n+1/2}}{2h_{y}}\label{eq:Fz}\\
\Phi_{x(i,j+1/2,k+1/2)}^{n+1/2} & = & -\left(\frac{\partial E_{z}}{\partial y}-\frac{\partial E_{y}}{\partial z}\right)|_{i,j+1/2,k+1/2}^{n+1/2}=\label{eq:Fx}\\
& - & \frac{E_{z(i,,j+1,k+1/2)}^{n+1/2}-E_{z(i,,j,k+1/2)}^{n+1/2}}{h_{y}}+\frac{E_{y(i,,j+1/2,k+1)}^{n+1/2}-E_{y(i,,j-1/2,k+1/2)}^{n+1/2}}{h_{z}}\nonumber 
\end {eqnarray}

Similar formulas are obtained~for the fields at the half-time steps, where all fields and currents are shifted by a half time step.

The use of the transport vectors $\mathbf{T}_{\perp}$ makes the boundary conditions in the $x-$direction trivial. One sets the inbound T vectors equal to the incident laser pulse and outbound T vectors to zero at the boundaries. This procedure absorbs waves normally incident on the boundaries {\it exactly}. 

It seems that we have to maintain two sets of fields for each time
step: fields at the full step and at the half step. The particles
however, can be pushed just once per time step. For the particle push
we use the symplectic semi-implicit mid-point scheme of Higuera and Hary \cite{Higuera} (pushers of Boris \cite{Boris} and Vay \cite{VayPusher} produce hardly discernible results) at the full time step: 

\begin{equation}
\mathbf{p}_{\alpha}^{n+1/2}=\mathbf{p}_{\alpha}^{n-1/2}+\tau q\left(\mathbf{E}^{n}+\frac{1}{\gamma m c}\mathbf{p} \times \mathbf{B}^{n}\right)\label{eq:push}
\end{equation}
where $\mathbf{p}=(\mathbf{p}_{\alpha}^{n+1/2}+\mathbf{p}_{\alpha}^{n-1/2})/2$ and
 $\gamma = \sqrt{1+p^2/m^2c^2}$.

These momenta are used to generate currents $\mathbf{j}^{n+1/2}$
at the half time steps. Currents at the full time step required to
push the half-time step fields can be obtained by simple averaging
on the grid

\begin{equation}
\mathbf{j}^{n}=\frac{1}{2}\left(\mathbf{j}^{n-1/2}+\mathbf{j}^{n+1/2}\right)
\end{equation}
To ensure the Lorentz invariance and charge conservation of the scheme, the current components
are defined within the cell at the same positions as the corresponding
$\mathbf{E}-$field components.

\section{Conservation laws on the RIP grid}

The RIP scheme places fields in a transverse plane as seen in Fig.\ref{fig:RIP}.
These field locations are perfectly suited for conservative definition
of the currents, charges, field divergence and curl on the grid. The
simple rule is that the trapezoidal formula must be applied in the
longitudinal direction, while in the transverse direction, the usual
Yee (spatial leap-frog) formula remains valid.

\subsection{Generalized rigorous charge conservation}

The numerical
continuity equation on the RIP grid has the form

\begin {eqnarray}
 & &c\frac{\rho_{i+1/2.j,k}^{n+1}-\rho_{i+1/2.j,k}^{n}}{\Delta} =  -\frac{1}{\Delta}\left(j_{x(i+1,j,k)}^{n+1/2}-j_{x(i,j,k)}^{n+1/2}\right)\label{eq:Continuity}\\
& - & \frac{1}{2h_{y}}\left(j_{y(i+1,j+1/2,k)}^{n+1/2}+j_{y(i,j+1/2,k)}^{n+1/2}-j_{y(i+1,j-1/2,k)}^{n+1/2}-j_{y(i,j-1/2,k)}^{n+1/2}\right)\nonumber \\
& - & \frac{1}{2h_{z}}\left(j_{z(i+1,j,k+1/2)}^{n+1/2}+j_{z(i,j,k+1/2)}^{n+1/2}-j_{z(i+1,j,k-1/2)}^{n+1/2}-j_{z(i,j,k-1/2)}^{n+1/2}\right)\nonumber 
\end {eqnarray}

The charge conservation on the grid can be enforced in various ways. One can correct currents and solve an elliptic problem after each time step \cite{Birdsall=000026Langdon}. One can atomatically calculate currents inside the cell  co that the charge is locally conserved \cite{Esirkepov}. One can move particles in a zig-zag along the axises \cite{Umeda2003}. The charge-conserving closure is not unique and an infinite number of other schemes can be easily generated like charge splitting curvy trajectory particle motion, etc. All these schemes generate different curl currents on the grid and thus have different noise properties.

However, the only true 2-nd order accurate current closure is the rigorous charge conservation method introduced originally by Villacenor and Buneman \cite{Villacenor}. This scheme assumes the straight particle trajectory during the time step. All other methods fail to do so. The Esirkepov scheme \cite{Esirkepov} coincides with the  Villacenor and Buneman  scheme {\it identically } as long as the particle stays inside one cell during the time step. It gives different results, however, as soon as the particle crosses boundaries.

We use a generalized rigorous charge conservation (GRCC) method based on \cite{Villacenor}, compare also \cite{Eastwood}.  It is not limited to the Cartesian geometry and is valid for any particle shape. Let us suppose, we have selected a form-factor $\mathbf{w}$ for the current deposition by the numerical macro-particles. The macro-particle $\alpha$ will then induce an instantaneous current $\mathbf{j(t)}$ on the grid with components:

\begin {eqnarray}
j_{x(i,j,k)}(t) & = & w^\alpha_x (\mathbf{r}^\alpha(t)) \label{eq:Current}\\
j_{y(i,j+1/2,k)}(t) & = & w^\alpha_y (\mathbf{r}^\alpha(t)) \nonumber \\
j_{z(i,j,k+1/2)}(t) & = & w^\alpha_z (\mathbf{r}^\alpha(t))  \nonumber 
\end {eqnarray}

\noindent where $\mathbf{r}^\alpha(t)$ is the instantaneous particle position inside the cell. Depending on the form-factor, the particle may induce instantaneous currents at many grid cells. We write expressions for one cell only, as the others are analogous.

The particle starts its motion at the time step $t^n$ at the position  $\mathbf{r}^\alpha(t^n)$ and finishes at the time step $t^{n+1} = t^n + \Delta/c$ at the position  $\mathbf{r}^\alpha(t^{n+1})$. The straight particle trajectory is parameterized as 
 $\mathbf{r}^\alpha(t) =  \mathbf{r}^\alpha(t^n) +   (t-t^n)/(t^{n+1} - t^{n})\mathbf{\delta r}^\alpha$, where  $\mathbf{\delta r}^\alpha = \mathbf{r}^\alpha(t^{n+1}) - \mathbf{r}^\alpha(t^{n})$. The current induced by the particle on the grid during the time step is then

\begin {equation}
\mathbf{j}^{n+1/2} = \frac{c}{\Delta} \int_0^1 \mathbf{w^\alpha} (\mathbf{r}^\alpha(t^n) +\tau \mathbf{\delta r}^\alpha ) d\tau = \mathbf{W^\alpha} (\mathbf{r}^\alpha(t^n), \mathbf{\delta r}^\alpha)  \label{eq:GRCC}
\end {equation}

\noindent When the macroparticle crosses the cell boundaries, the integral (\ref{eq:GRCC}) must be split  along the straight particle trajectory in  parts, where the particle center belongs to one particular cell. 

For most popular particle shapes (box, triangle, quadratic, spline, etc.), the integration in (\ref{eq:GRCC}) is done analytically (using  any  symbolic integration software) and the function $\mathbf{W^\alpha} (\mathbf{r}^\alpha(t^n), \mathbf{\delta r}^\alpha)$ is easily coded. Villacenor and Buneman did it explicitly for the case of Cartesian grid and a square box particle shape.

This described GRCC algorithm preserves the discretized Maxwell-Gauss equations automatically  by the RIP scheme.

\subsection{Divergence conservation}

In the same manner we define the curl and divergence of the fields.
For example, $\nabla \cdot \mathbf{B}$  is defined as 

\begin {eqnarray}
& & \nabla \cdot \mathbf{B}_{i+1/2,j+1/2,k+1/2}  =  \frac{B_{x(i+1,j+1/2,k+1/2)}-B_{x(i-1,j+1/2,k+1/2)}}{\Delta}\label{eq:divB}\\
& + & \frac{B_{y(i+1,j+1,k+1/2)} + B_{y(i,j+1,k+1/2)} -B_{y(i+1,j,k+1/2)} - B_{y(i,j,k+1/2)}}{2h_y}\nonumber \\
& + & \frac{B_{z(i+1,j+1/2,k+1)} + B_{z(i,j+1/2,k+1)} -B_{z(i+1,j+1/2,k)} - B_{z(i,j+1/2,k)}}{2h_z}\nonumber 
\end {eqnarray}

\noindent at the middle cell position $\left(i+1/2,j+1/2,k+1/2\right)$. We average
fields along the $X-$axis according to the trapezoidal rule while
the usual leap-frog Yee rule is applied along the transverse coordinates. 
It is straightforward to check that the RIP scheme preserves $\nabla \cdot \mathbf{B}$ defined in this way.

The fields at the half-time steps are required to calculate the diffraction
terms only. Without diffraction, the need to maintain the additional
set of fields at half-time steps vanishes and the RIP scheme becomes
identical to the standard 1D PIC scheme \cite{Birdsall=000026Langdon},
which is the workhorse of 1D plasma simulations due to its excellent
stability and accuracy.

\begin{figure}
\includegraphics[scale=0.5]{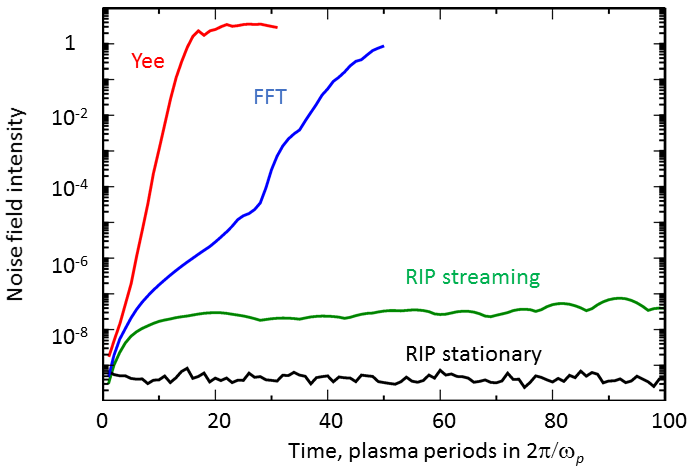}\caption{ (color online)  Intensity of fluctuating fields in the ``streaming plasma'' simulations.
The Yee scheme is fully subject to the numerical Cerenkov instability
and reaches saturation within a few plasma periods. The FFT-based
solver avoids the first order numerical Cerenkov resonance and is
subject to second order aliasing resonance. The RIP simulation of
streaming plasma shows several orders of magnitude lower noise fields.
The noise field growth rate is very low here. Mention that the FFT
solver (\ref{eq:FFT}) \textit{is identical} to the RIP solver for
waves running along the $X-$axis. Yet, the FFT solver is subject
to NCI because of aliasing. The stationary plasma case shows no instability
at all in the RIP simulation. \label{fig:Cerenkov}\label{fig:RIP-1}}
\end{figure}
\begin{verbatim}

\end{verbatim}

\section{Dispersion and stability of the RIP scheme}

We apply the plane-wave analysis to the marching equations (\ref{eq:Ex}),
(\ref{eq:Bx}) and (\ref{eq:Ey+march})-(\ref{eq:Bz+march}) with
the refraction/diffraction terms (\ref{eq:Gy})-(\ref{eq:Fx}) assuming
$\mathbf{F}=\mathbf{\tilde{F}}\exp\left(-i\omega t+i\mathbf{k}\mathbf{r}\right)$.
For simplicity, we assume uniform plasma frequency $\omega_{p}^{2}=4\pi ne^{2}/\gamma$
and the linear current response to the electric field $\frac{2c}{\Delta}\sin\frac{\omega\tau}{2}\tilde{\mathbf{J}}=iq^{2}n\tilde{\mathbf{E}}$.
For the case of interest, $c\tau=h_{x}=\Delta$, these equations become

\begin {eqnarray}
\frac{2}{\Delta}\sin\frac{\omega\Delta}{2c}\cos\frac{k_{x}\Delta}{2}\tilde{E}_{y} & =& -\frac{2}{\Delta}\sin\frac{k_{x}\Delta}{2}\cos\frac{\omega\Delta}{2c}\tilde{B}_{z}\label{eq:Ey-1}\\
 & + & \frac{2}{h_{z}}\sin\frac{k_{z}h_{z}}{2}\cos\frac{k_{x}\Delta}{2}\tilde{B}_{x}+\omega_{p}^{2}\frac{\Delta}{2c\sin\frac{\omega\Delta}{2c}}\tilde{E}_{y}\nonumber \\
\frac{2}{\Delta}\sin\frac{\omega\Delta}{2c}\cos\frac{k_{x}\Delta}{2}\tilde{E}_{z} & = & \frac{2}{\Delta}\sin\frac{k_{x}\Delta}{2}\cos\frac{\omega\Delta}{2c}\tilde{B}_{y}\label{eq:Ez-1}\\
 & - & \frac{2}{h_{y}}\sin\frac{k_{y}h_{y}}{2}\cos\frac{k_{x}\Delta}{2}\tilde{B}_{x}+\omega_{p}^{2}\frac{\tau}{2\sin\frac{\omega\tau}{2}}\tilde{E}_{z}\nonumber \\
\frac{2}{\Delta}\sin\frac{\omega\Delta}{2c}\tilde{E}_{x} & = & \frac{2}{h_{y}}\sin\frac{k_{y}h_{y}}{2}\tilde{B}_{z}-\frac{2}{h_{z}}\sin\frac{k_{z}h_{z}}{2}\tilde{B}_{y}\label{eq:Ex-1}\\
 & + & \omega_{p}^{2}\frac{\Delta}{2c\sin\frac{\omega\Delta}{2c}}\tilde{E}_{x}\nonumber 
\end {eqnarray}

\begin {eqnarray}
\frac{2}{\Delta}\sin\frac{\omega\Delta}{2c}\cos\frac{k_{x}\Delta}{2}\tilde{B}_{y} & = & \frac{2}{\Delta}\sin\frac{k_{x}\Delta}{2}\cos\frac{\omega\Delta}{2c}\tilde{E}_{z}\label{eq:By-1}\\
 & - & \frac{2}{h_{z}}\sin\frac{k_{z}h_{z}}{2}\cos\frac{k_{x}h_{x}}{2}\tilde{E}_{x}\nonumber \\
\frac{2}{\Delta}\sin\frac{\omega\Delta}{2c}\cos\frac{k_{x}\Delta}{2}\tilde{B}_{z} & = & -\frac{2}{\Delta}\sin\frac{k_{x}\Delta}{2}\cos\frac{\omega\Delta}{2c}\tilde{E}_{y}\label{eq:Bz-1}\\
 & + & \frac{2}{h_{y}}\sin\frac{k_{y}h_{y}}{2}\cos\frac{k_{x}h_{x}}{2}\tilde{E}_{x}\nonumber \\
\frac{2}{\Delta}\sin\frac{\omega\tau}{2}\tilde{B}_{x} & = & \frac{2}{k_{y}h_{y}}\sin\frac{k_{y}h_{y}}{2}\tilde{E}_{z}-\frac{2}{k_{z}h_{z}}\sin\frac{k_{z}h_{z}}{2}\tilde{E}_{y}\label{eq:Bx-1}
\end {eqnarray}

The dispersion relation in vacuum ($\omega_{p}=0)$ is rather simple: 

\begin{eqnarray}
\left(\frac{1}{h_{y}^{2}}\sin^{2}\frac{k_{y}h_{y}}{2}+\frac{1}{h_{z}^{2}}\sin^{2}\frac{k_{z}h_{z}}{2}\right) & + & \frac{1}{\Delta^{2}}\sin^{2}\frac{\Delta k_{x}}{2}\left(1-\Delta^{2}\left(\frac{1}{h_{y}^{2}}\sin^{2}\frac{k_{y}h_{y}}{2}+\frac{1}{h_{z}^{2}}\sin^{2}\frac{k_{z}h_{z}}{2}\right)\right) \nonumber \\ 
& = & \frac{1}{\Delta^{2}}\sin^{2}\frac{\Delta\omega}{2c}\label{eq:Dispersion}
\end{eqnarray}
The stability condition in vacuum is

\begin{equation}
\Delta^{2}\left(\frac{1}{h_{y}^{2}}+\frac{1}{h_{z}^{2}}\right)<1
\end{equation}
In the presence of plasmas, it is modified to

\begin{equation}
\frac{1}{\Delta^{2}}>\frac{1}{h_{y}^{2}}+\frac{1}{h_{z}^{2}}+\frac{\omega_{p}^{2}}{4c^{2}}
\end{equation}
The RIP scheme combines dispersionless properties of the standard
1D solver along the $X-$axis with the Yee dispersion for waves running
in the transverse direction. Indeed, setting $k_{y}=k_{z}=0$ in the
dispersion relation (\ref{eq:Dispersion}), we immediately obtain
$\omega=ck_{x}$ and the phase velocity

\begin{equation}
V_{ph}=\frac{\omega}{k_{x}}=c
\end{equation}
 for plane waves propagating in the $X-$direction.

Conversely, setting $k_{x}=0$, we obtain the usual 2D Yee dispersion
relation for waves propagating in the transverse direction

\begin{equation}
\frac{1}{h_{y}^{2}}\sin^{2}\frac{k_{y}h_{y}}{2}+\frac{1}{h_{z}^{2}}\sin^{2}\frac{k_{z}h_{z}}{2}=\frac{1}{\Delta^{2}}\sin^{2}\frac{\Delta\omega}{2c}\label{eq:Dispersion-1}
\end{equation}
with all its known advantages and drawbacks.

Mention that the NDF scheme introduced in \cite{VLPL1999} has a different stability condition:

\begin{equation}
\frac{1}{c^2 \tau^{2}}>\frac{1}{h_{x}^{2}}+\frac{\omega_{p}^{2}}{4c^{2}}
\end{equation}

\noindent so that one can set $c\tau = h_x$ only in vacuum. The presence of plasma, $\omega_p > 0$, one has to choose $c\tau < h_x$ and the dispersionless properties of the NDF scheme are compromised.

\section{Numerical tests of the RIP Maxwel solver}

\subsection{Numerical Cerenkov instability test}

As a first test, we take the numerical Cerenkov instability. We compare
the standard Yee solver, the FFT-based solver (\ref{eq:FFT}) and
the RIP solver, all implemented on the VLPL platform \cite{VLPL_CERN}.
No artificial filtering of fields or currents is used. The initial
configuration is a ellipsoidal plasma of Gaussian density profile $n=n_{0}\exp\left(-r^{2}/\sigma^{2}\right)$
consisting of electrons and protons moving in the $X-$direction with
the average momentum $<\mathbf{p_{0}}>/m_{\alpha}c=\left(p_{x0},0,0\right)$,
where $\alpha$ denotes the particle type ($\alpha=e,p$), with $m_{p}/m_{e}=1846$.
To seed the instability, the electrons have a small initial temperature
$<\left(\mathbf{p}_{0}-<\mathbf{p}_{0}>\right)^{2}>=\sigma_{p}^{2}$.
In relativistically normalized units, the simulation parameters are:
the peak plasma density is $n_{0}=1$ with the corresponding non-relativistic
plasma frequency $\omega_{p}=\sqrt{4\pi n_{0}e^{2}/m_{e}}$. The initial
particle momenta $p_{x0}=-10$ and $\sigma_{p}=10^{-4}$. The grid
steps were $h_{y}=h_{z}=1.88\,c/\omega_{p}$ and $h_{x}=c\tau=\Delta=0.63\,c/\omega_{p}$.
As a diagnostics for the comparison, we selected the growth of the
maximum local field intensity $I=\mathbf{E^{2}}+\mathbf{B}^{2}$ on
the grid. The results are shown in Fig.\ref{fig:Cerenkov}

We see that the fluctuating fields in simulations using the Yee scheme
grow to the non-linear saturation within a few plasma oscillations.
This is because the Yee solver is exposed to the first order Cerenkov
resonance. The FFT-based solver is dispersionless and avoids the first
order Cerenkov resonance. Still, the second order aliasing of the
spectral FFT solver leads to the numerical Cerenkov instabilty, though
at a lower growth rate as compared with the standard Yee solver. 

\begin{figure}
\includegraphics[scale=0.5]{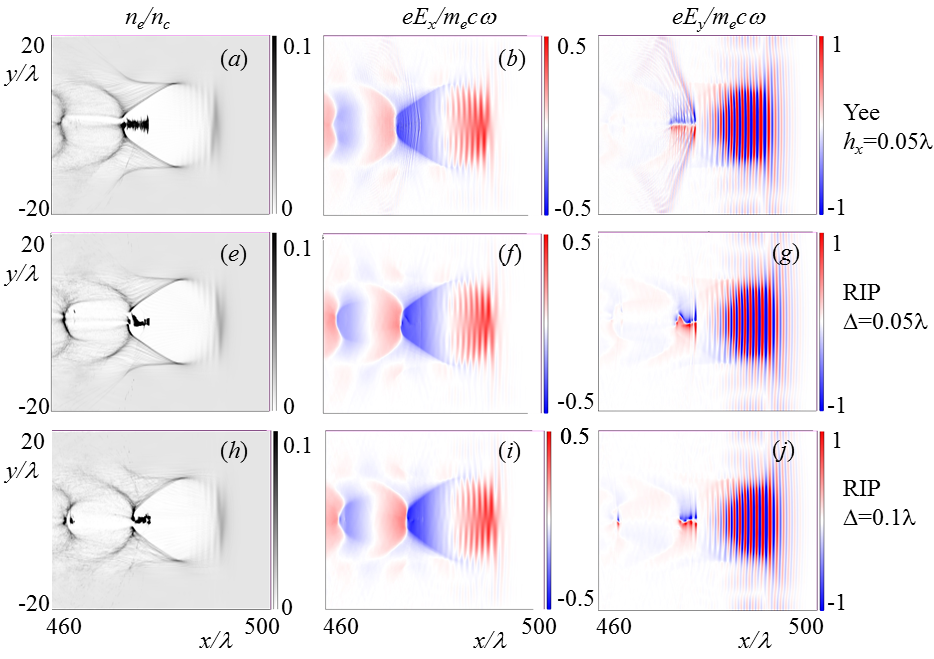}\caption{(color online) Laser-plasma wake field acceleration in the bubble regime. The electron
plasma density $n_{e}n_{c}$, the accelerating electric field $eE_{x}/mc\omega$
and the transverse electric field $eE_{y}/mc\omega$ are shown after
the laser pulse propagated $L_{a}=300\lambda$. The first row shows
the Yee scheme simulation results for the longitudinal grid step $h_{x}=0.05\lambda$
and time step $\tau=0.04\lambda/c$, the middle row gives the RIP
scheme results with the same grid steps and time step $\tau=h_{x}/c$,
and the last row shows RIP scheme results with two times rougher resolution
in the propagation direction $h_{x}=c\tau=0.1\lambda$. The numerical
Cerenkov resonance in the Yee scheme is clearly seen in frame (c)
as the short wavelength bow-like emission by the accelerated electron
bunch. \label{fig:Bubble}}
\end{figure}

In contrast, the RIP solver is free from NCI. The noise in the RIP
scheme remains many orders of magnitude lower over a long simulation
time of $t=100\cdot2\pi/\omega_{p}$. The very slow growth of the
noise fields here has nothing to do with the Cerenkov resonance, but
is the unavoidable ``numerical heating'' always present in PIC codes. 

\begin{figure}
\includegraphics[scale=0.7]{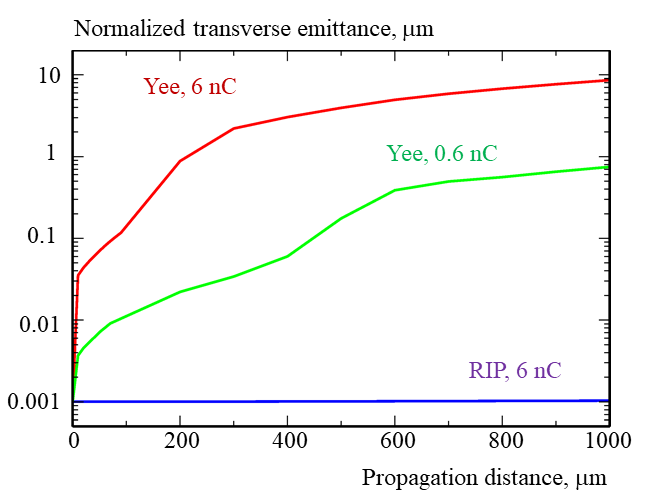}\caption{(color online) Ultra-low transverse emittance beam propagation in vacuum. The bunches carry charges of $6~$nC or $0.6$nC. The initial normalized transverse emittance is $\varepsilon_y = \varepsilon_z = 1~$nm. The Yee solver shows fast emittance growth. The RIP solver preserves the emittance at sub-nm level.
 \label{fig:emitEvolution}}
\end{figure}

Finally, we do another simulation with a stationary plasma, $<\mathbf{p_{0}}>/m_{\alpha}c=\left(0,0,0\right)$,
while keeping all other parameters the same. We observe here no numerical
heating at all. Intensity of fluctuating noise fields remains constant
over many hundreds of plasma periods here. The higher absolute level
of the noise for the streaming plasma is the natural consequence of
the larger initial noise current source in this case. Fig.\ref{fig:Cerenkov}
demonstrates clearly that the RIP scheme is much less subject to Cerenkov instability
for plasmas drifting along the selected axis.

We stress here that the FFT-based method (\ref{eq:FFT}) is \textit{identical}
to the RIP solver for waves running in the $X-$direction. Yet, we
observe a quite different behaviour with respect to the numerical
Cerenkov instability. The reason for this difference has to be studied further.

We mention here that the numerical Cerenkov instability of uniformly
streaming plasma can be alleviated by using a co-moving grid as proposed
by Lehe et al. \cite{Lehe_Lagrangian}. The method exploits a Galilean
transformation to a grid in which the background plasma does not stream
through the cell boundaries. Yet, the dense bunch of accelerated particles
moves in the opposite direction at twice the light speed relative
to this grid and is fully exposed to the Cerenkov resonance.

\subsection{Laser-driven plasma bubble}

As the second numerical test, we select laser-plasma particle acceleration
in the bubble regime \cite{Bubble}. A circularly polarized laser
pulse with initial vector potential $\mathbf{A}=\Re\left[a(\xi,\mathbf{r_{\perp}})(\mathbf{e_{y}}+i\mathbf{e_{z}})\exp(ik\xi)\right]$
is used. Here, $\xi=x-ct$ and the envelope shape has been selected
as a ellipsoidal Gaussian $a(\xi,\mathbf{r_{\perp}})=a_{0}\exp(-\xi^{2}/\sigma_{||}^{2}-r_{\perp}^{2}/\sigma_{\perp}^{2})$
with the amplitude $a_{0}=5$ , length $\sigma_{||}=5\lambda$ and
radius $\sigma_{\perp}=5\lambda$, where the laser wavelength $\lambda=2\pi/k$.
The plasma consisting of electrons and protons has an initial density
$n=0.01\,n_{c}$, where $n_{c}=m_{e}\omega^{2}/4\pi e^{2}$ is the
critical density. At the plasma boundary, the density increases lineraly
from $n=0$ to $n=0.01\,n_{c}$ over a length $L=38\,\lambda$. The
simulation results after an acceleration distance of $L_{a}=300\,\lambda$
are shown in Fig.\ref{fig:Bubble}. The simulation box has the size
$40\lambda\times40\lambda\times40\lambda$. The grid steps are $h_{x}=0.05\lambda$,
$h_{y}=0.25\lambda$, $h_{z}=0.25\lambda$ and the time step is $\tau=0.045\lambda/c$
in the Yee simulation, and $\tau=h_{x}/c$ in the RIP simulation. 

\begin{figure}
\includegraphics[scale=0.6]{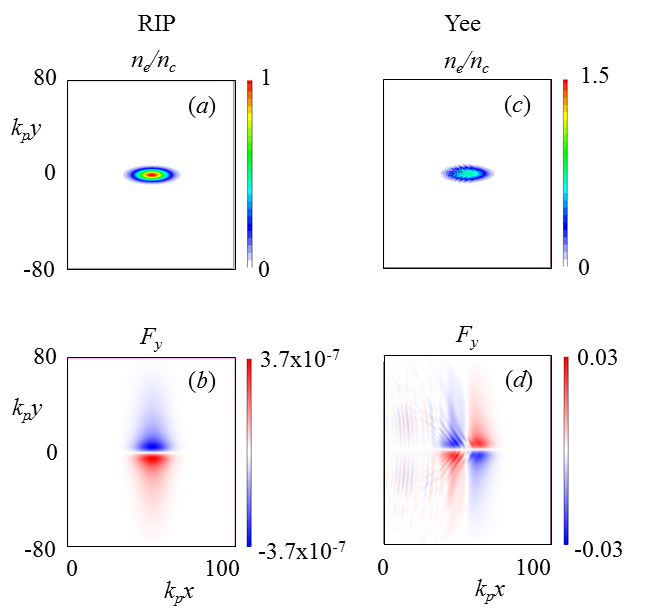}\caption{(color online) Beam propagation in vacuum. The bunch carries the charge of $6~$nC. The initial normalized transverse emittance is $\varepsilon_y = \varepsilon_z = 1~$nm. The frames show the bunch density $n/n_c$ and the combine transverse force $F_y = e(E_y-B_z)/mc\omega$. Here, $\omega=2\pi c/\lambda$, $\lambda=1~\mu$m and $n_c = m \omega^2/4\pi e^2$. The frames (a) and (b) are simulated with the RIP solver and taken after $1~$mm propagation distance. The frames (c) and (d) are simulated with the Yee solver and are taken after $0.1~$mm propagation. The RIP solver shows the correct transverse force due to the bunch self-action. The Yee solver is spoiled by poor force interpolation and the onset of numerical Cerenkov instability.
 \label{fig:emit}}
\end{figure}

We see that the trapped electron bunch of the bubble has a fine longitudinal
structure in the Yee simulation. At the same time, the bubble accelerating
field $E_{x}$ is rippled with the short-wavelength radiation emitted
by the relativistic electrons due to the numerical Cerenkov resonance.
This numerical emission is clearly seen in Fig.~\ref{fig:Bubble}(c)
as the bow-like short wavelength radiation emanating from the dense
electron bunch. The RIP simulation shows a rather smooth electron
bunch and no signatures of numerical Cerenkov emission. The $E_{y}-$field
of the relativistic electron bunch has a clean quasi-static form:
it is not bow-shaped, but perpendicular to the bunch. Further, a small
additional numerical dephasing can be observed at the leading edge
of the bubble.

To check the RIP scheme convergence, we did an additional simulation
with rough resolution. We doubled the longitudinal grid step and the
time step to $h_{x}=c\tau=0.1\lambda$, so that we have only 10 cells
per laser wavelength. The results are shown in the last row in Fig.~\ref{fig:Bubble}.
One observes little difference from the higher resolution simulation,
shown in the middle row in Fig.~\ref{fig:Bubble}. Compare the phase of the laser pulse seen in the electron density perturbations in frames ({\it f}) and ({\it i}). 
The RIP simulation even at this rough resolution accurately describes the laser phase. The Yee solver gives a completely wrong laser phase as seen in the frame ({\it b}). 
At the resolution of just 10 cells per wavelength (or even smaller), the RIP solver is a good alternative to full electromagnetic PIC codes that employ the envelope approximation \cite{INFERNO,CowanEnvelope} when the laser pulse is short.

\subsection{Ultra-low transverse emittance beam propagation in vacuum}

Finally, we check how well the RIP scheme preserves the beam emittance. The future colliders and XFEL light sources must have beams 
with ultra-low transverse emittance. Emittances in the range of a few nanometers, or even picometers have to be realized. 
There are several approaches, how such beams can be generated using the conventional accelerators.
Plasma-based acceleration also might reach such ultra-low beam emittances, using, e.g. the Trojan-horse injection \cite{Trojan}.
Thus, very accurate simulation methods are required, where the emittance is preserved at picometer levels.

Lehe et al \cite{Lehe_Emittance} have shown that the Yee Maxwell solver has problems with emittance conservation and suggested 
an "improved" one that shows a better emittance preservation. There are several sources of emittance growth in PIC simulations: 
numerical heating, numerical Cerenkov instability and the wrong field interpolation to the particle position due the staggered mesh. 
Lehe et al. modified the Yee solver to remove the zero-order Cerenkov resonance. This improved the emittance preservation \cite{Lehe_Emittance}.

In this sub-section, we simulate an electron bunch propagation in vacuum. The electron bunch has initial energy of 10 GeV and normalized transverse emittance of
$\varepsilon_y = \varepsilon_z = 1~$nm. The bunch has a Gaussian shape with $\sigma_x = \sigma_y = \sigma_z = 1.41~\mu$m. We simulate two cases, where the bunch carries either $0.6~$nC or $6.0~$nC charge.
The simulations are done using the Yee or RIP solver for total propagation distance of $1~$mm. The grid steps were $h_x = 0.1~\mu$m, $h_y = h_z = 0.15~\mu$m. The time step for the Yee solver was $c\tau = 0.7 h_x$.

The emittance evolution is shown in Fig.~\ref{fig:emitEvolution}. The Yee solver shows a very fast inital jump of the bunch emittance due to the incorrect field interpolation to the particle positions on the staggered Yee mesh. The jump is higher for the higher bunch charge. Later, the bunch becomes unstable due to the numerical Cerenkov instability and the emittance grows steadily. After the full propagation distance of $1~$mm, the final numerical emittance grew in the Yee simulations to $8.5~\mu$m for the high current bunch of $6~$nC and to $0.8~\mu$m for the low current bunch of $0.6~$nC. 

The RIP solver shows an excellent conservation of the emittance. For the high-current bunch of $6~$nC, the transverse emittance grew only by $10~$pm, from $1~nm$ to $1.01~$nm after the full $1~mm$ propagation distance. 
This miniscule emittance growth  is physical: it is due to the Coulomb explosion of the high current bunch.

The bunch dynamics is shown in Fig.~\ref{fig:emit}. It shows the normalized bunch density $n_e/n_c$ and the tranverse force acting on the electrons $F_y = e(E_y-B_z)/mc\omega$. Here, $\omega=2\pi c/\lambda$, $\lambda=1~\mu$m and $n_c = m \omega^2/4\pi e^2$. The frames (a) and (b) are taken for the RIP solver after the full $1~$mm propagation distance, while the frames (c) and (d) are taken for the Yee solver after $0.1~$mm propagation. We take here the high current case of $6~$nC charge bunch. 

The Yee solver fails with the transverse force by many orders of magnitude. We also see the development of NCI in the transverse force causing self-modulation of the bunch tail. Apparently, the Yee solver is not the best choice when one wants to simulate low emittance bunches. 

The RIP solver accurately reproduces the transverse force due to the self-interaction down to the machine precision. The normalized transverse force is at the level of $F_y = 3.7\cdot 10^{-7}$ in this case. Thus, the RIP Maxwell solver is perfectly suited to simulate bunches with sub-nanometer emittances.

\section{Discussion}

The new RIP scheme is a compact stencil FDTD Maxwell solver that removes
the numerical dipersion in one selected direction. For the waves propagating
in the transverse direction, it corresponds to the Yee solver. The
RIP scheme is local and does not use any global spectral method. This
allows for efficient parallelization via domain decomposition in all
three dimensions. The computational costs of the RIP solver is comparable
with that of the standard Yee solver. The RIP solver can be used for
simulations of quasi-1D physics problems like laser wake field acceleration.
This RIP marching algorithm has a form of ``Galiliean field shift''
exactly by single cell per time step. Thus, instead of shifting the
grid following the relativistic plasma \cite{Lehe_Lagrangian}, the
RIP solver shifts the transverse fields so that the relativistic particle
sees the same fields when it enters the new cell. Apparently, this
procedure greatly reduces the numerical Cerenkov instability.

\section*{Acknowledgements}

This work has been supported in parts by BMBF and DFG (Germany).



\section*{References}

\begin{thebibliography}{00}

\bibitem{ALEGRO}ALEGRO Collaboration, ``Towards an Advanced Linear
International Collider'' arXiv:1901.10370 (2019)

\bibitem{Leemans2009}W. Leemans and E. Esarey, ``Laser-driven plasma-wave
electron accelerators'' Physics Today \textbf{62}, 3, 44 (2009)

\bibitem{EUpraxia} http://eupraxia-project.eu

\bibitem{Joshi_Caldwell}C. Joshi and A. Caldwell, ``Plasma Accelerators'',
in Accelerators and Colliders, edited by S. Myers and H. Schopper
(Springer Berlin Heidelberg, Berlin, Heidelberg, 2013), Chap. 12.1,
pp. 592 \textendash{} 605.

\bibitem{Tajima_Dawson}T. Tajima and J. M. Dawson. Phys. Rev. Lett.
\textbf{43}, 267 (1979)

\bibitem{Esarey_review}E. Esarey, C. B. Schroeder, W. P. Leemans,
Rev. Mod. Phys. \textbf{81}, 1229 (2009).

\bibitem{OSIRIS}Fonseca RA, Martins SF, Silva LO, Tonge JW, Tsung
FS, Mori WB, ``One-to-one direct modeling of experiments and astrophysical
scenarios: pushing the envelope on kinetic plasma simulations'',
Plasma Physics and Controlled Fusion vol. \textbf{50}, 124034 (2008)

\bibitem{VLPL_CERN}A. Pukhov, ``Particle-In-Cell Codes for Plasma-based
Particle Acceleration'', CERN Yellow Rep. 1 , 181 (2016).

\bibitem{EPOCH}T. D. Arber, K. Bennett, C. S. Brady, A. Lawrence-Douglas,
M. G. Ramsay, N. J. Sircombe, P. Gillies, R. G. Evans, H. Schmitz,
A. R. Bell and C. P. Ridgers. ``Contemporary particle-in-cell approach
to laser-plasma modelling'', Plasma Phys. Control. Fusion \textbf{57},
113001 (2015).

\bibitem{Vay}Jean-Luc Vay, Irving Haber, Brendan B. Godfrey, ``A
domain decomposition method for pseudo-spectral electromagnetic simulations
of plasmas'', Journal of Computational Physics \textbf{243}, 260
(2013).

\bibitem{Smilei}J. Derouillat, A. Beck, F. Perez, T. Vinci, M. Chiaramello,
A. Grassi, M. Fle, G. Bouchard, I. Plotnikov, N. Aunai, J. Dargent,
C. Riconda, M. Grech, SMILEI: a collaborative, open-source, multi-purpose
particle-in-cell code for plasma simulation, Comput. Phys. Commun.
\textbf{222}, 351-373 (2018)

\bibitem{PIConGPU}Heiko Burau, Renee Widera, Wolfgang Hönig, Guido Juckeland, Alexander Debus, Thomas Kluge, and M. Bussmann  "PIConGPU: A Fully Relativistic Particle-in-Cell Code for a GPU Cluster", IEEE Transactions on Plasma Science \textbf{38}, 2831-2839, (2010)

\bibitem{wikiPIC} https://en.wikipedia.org/wiki/Particle-in-cell

\bibitem{Birdsall=000026Langdon}C.K. Birdsall and A.B. Langdon, Plasma
Physics via Computer Simulations (Adam Hilger, New York, 1991). http://dx.doi.org/10.1887/0750301171

\bibitem{Leemans_8_GeV}A. ~ J. Gonsalves, K. Nakamura,
J. Daniels, C. Benedetti, C. Pieronek, T. ~ C. ~ H.
de Raadt, S. Steinke, J. ~ H. Bin, S. ~ S. Bulanov,
J. van Tilborg, C. ~ G. ~ R. Geddes, C. ~ B.
Schroeder, Cs. Toth, E. Esarey, K. Swanson, L. Fan-Chiang, G. Bagdasarov,
N. Bobrova, V. Gasilov, G. Korn, P. Sasorov, and W. ~ P.
Leemans ``Petawatt Laser Guiding and Electron Beam Acceleration to
8 GeV in a Laser-Heated Capillary Discharge Waveguide'' Phys. Rev.
Lett. \textbf{122}, 084801

\bibitem{SALC}I. Blumenfeld, C. E. Clayton, F.-J. Decker, M. J. Hogan,
C. Huang, R. Ischebeck, R. Iverson, C. Joshi, T. Katsouleas, N. Kirby,
W. Lu, K. A. Marsh, W. B. Mori, P. Muggli, E. Oz, R. H. Siemann, D.
Walz, and M. Zhou, ``Energy doubling of 42 ~ GeV electrons
in a metre-scale plasma wakefield accelerator'' Nature (London) \textbf{445}
, 741 (2007).

\bibitem{AWAKE}E. Adli et al., AWAKE collaboration ``Acceleration
of electrons in the plasma wakefield of a proton bunch'', Nature
\textbf{561} (7723), 363 (2018)

\bibitem{QUICKPIC}C. Huang, V.K. Decyk, C. Ren, M. Zhou, W. Lu, W.B.
Mori, J.H. Cooley, T.M. Antonsen, T. Katsouleas ``QUICKPIC: A highly
efficient particle-in-cell code for modeling wakefield acceleration
in plasmas'' Journal of Computational Physics \textbf{217} (2006)
658\textendash 679

\bibitem{LCODE}K. V. Lotov, ``Fine wakefield structure in the blowout
regime of plasma wakefield accelerators'' Phys. Rev. ST Accel. Beams
\textbf{6}, 061301 (2003).

\bibitem{WAKE}P. Mora and Th. Antonsen Jr ``Kinetic modeling of
intense, short laser pulses propagating in tenuous plasmas'' Phys.
Plasmas, \textbf{4}, 217 (1997).

\bibitem{Vay_Boost}J.-L. Vay ``Noninvariance of Space- and Time-Scale
Ranges under a Lorentz Transformation and the Implications for the
Study of Relativistic Interactions'' Phys. Rev. Lett. \textbf{98},
130405 (2007)


\bibitem{FBPIC} Remi Lehe. Manuel Kirchen, Igor A.Andriyash. Brendan B.Godfrey, Jean-Luc Vay "A spectral, quasi-cylindrical and dispersion-free Particle-In-Cell algorithm'' Comp. Phys. Comm. \textbf{203},
66-82 (2016)  https://doi.org/10.1016/j.cpc.2016.02.007

\bibitem{INFERNO}C. Benedetti, C.B. Schroeder, E. Esarey, W.P. Leemans "Efficient modeling of laser-plasma
accelerator staging experiments using
INF\&RNO"  AIP Conference Proceedings 1812, 050005 (2017); https://doi.org/10.1063/1.4975866

\bibitem{Cerenkov_in_PIC1974}B. Godfrey, Numerical Cherenkov instabilities
in electromagnetic particle codes, Journal of Computational Physics
15 (4) (1974) 504\textendash 521.

\bibitem{Yee}Yee K S ``Numerical solution of initial boundary value
problems involving maxwell's equations in isotropic media'' 1966
IEEE Trans. Antennas Propag. 14 302\textendash 7

\bibitem{Vay_Cerenkov_Mitigation}J.-L. Vay, C.G.R. Geddes, E. Cormier-Michel,
D.P. Grote ``Numerical methods for instability mitigation in the
modeling of laser wakefield accelerators in a Lorentz-boosted frame''
Journal of Computational Physics \textbf{230} (2011) 5908\textendash 5929.

\bibitem{Relativistic_Beam_Cerenkov}D.-Y. Na, J. L. Nicolini, R.
Lee, B.-H. V. Borges, Y. A. Omelchenko, F. L. Teixeira ``Diagnosing
numerical Cherenkov instabilities in relativistic plasma simulations
based on general meshes'' arXiv:1809.05534 (2019)

\bibitem{Tixon} R. Nuter, V. Tikhonchuk, Suppressing the numerical
cherenkov radiation in the yee numerical scheme, Journal of Computational
Physics \textbf{305} (2016) 664 \textendash{} 676. doi:https://doi.org/10.1016/
j.jcp.2015.10.057.

\bibitem{Lehe_Emittance}R. Lehe, A. Lifschitz, C. Thaury, and V.
Malka ``Numerical growth of emittance in simulations of laser-wakefield
acceleration'', PHYSICAL REVIEW SPECIAL TOPICS - ACCELERATORS AND
BEAMS \textbf{16}, 021301 (2013)

\bibitem{Cowan}Benjamin M. Cowan, David L. Bruhwiler, John R. Cary,
and Estelle Cormier-Michel, Cameron G. R. Geddes ``Generalized algorithm
for control of numerical dispersion in explicit time-domain electromagnetic
simulations'' PHYSICAL REVIEW SPECIAL TOPICS - ACCELERATORS AND BEAMS
\textbf{16}, 041303 (2013)

\bibitem{Cole}J.B. Cole ``High-accuracy Yee algorithm based on nonstandard
finite differences: new developments and verifications'' IEEE Transactions
on Antennas and Propagation \textbf{50}, 1185 - 1191 (2002)

\bibitem{NPQED}V Yakimenko, S Meuren, F Del Gaudio, C Baumann, A
Fedotov, F Fiuza, T Grismayer, MJ Hogan, A Pukhov, LO Silva, G White,
``Prospect of Studying Nonperturbative QED with Beam-Beam Collisions'',
Phys. Rev. Lett. 122, 190404 (2019).

\bibitem{LeheLens} R. Lehe, C. Thaury, E. Guillaume, A. Lifschitz, and V. Malka "Laser-plasma lens for laser-wakefield accelerators"  Phys. Rev. ST Accel. Beams 17, 121301 (2014)

\bibitem{Haber}I. Haber, R. Lee, H. Klein, J. Boris, Advances in
electromagnetic simulation techniques, in: Proc. Sixth Conf. on Num.
Sim. Plasmas, Berkeley, CA, 1973, pp. 46\textendash 48.

\bibitem{Buneman_Tristan}Buneman, O., Barnes, Barnes,Barnes, C. W.,
Green, J. C., and Nielsen, D. E., Review: Principles and capabilities
of 3-d, E-M particle simulations, J. Comput. Phys., 38, 1, 1980.

\bibitem{Vay_convergence}P. Lee, J.-L. Vay ``Convergence in nonlinear
laser wakefield accelerators modeling in a Lorentz-boosted frame''
Computer Physics Communications 238 (2019) 102\textendash 110

\bibitem{Sommerfeld}Sommerfeld A., Vorlesungen uber Theoretische
Physik. Band 3: Elektrodynamik, Dieterichsche Verlagsbuchhandlung,
Wiesbaden 1948.



\bibitem{Vay2016} H. Vincenti and J.-L. Vay, \textquotedblleft Detailed analysis of the effects of stencil spatial variations with arbitrary high-order finite-difference Maxwell solver\textquotedblright , 
Computer Physics Communications 220,
147-167 (2016), https://doi.org/10.1016/j.cpc.2015.11.009

\bibitem{Ultra-High_MAxwell} H.Vincenti and J.-L.Vay \textquotedblleft Ultrahigh-order
Maxwell solver with extreme scalability for electromagnetic PIC simulations
of plasmas\textquotedblright , Computer Physics Communications 228,
22-29 (July 2018), https://doi.org/10.1016/j.cpc.2018.03.018

\bibitem{Aliasing_PIC}Brendan B. Godfrey ``Review and Recent Advances
in PIC Modeling of Relativistic Beams and Plasmas'' arxiv:1408.1146
(2014)

\bibitem{CerenkovFFT}B.B. Godfrey, J.-L. Vay, Comput. Phys. Comm.
\textbf{196} (2015) 221\textendash 225, http://dx. doi.org/10.1016/j.cpc.2015.06.008.

\bibitem{Cerenko_ Filter}Fei Li, Peicheng Yu, Xinlu Xu, Frederico
Fiuza, Viktor K. Decyk, Thamine Dalichaouch, Asher Davidson, Adam
Tableman, Weiming An, Frank S. Tsung, Ricardo A. Fonseca, Wei Lu,
Warren B. Mori ``Controlling the numerical Cerenkov instability in
PIC simulations using a customized finite difference Maxwell solver
and a local FFT based current correction'' Computer Physics Communications
214 (2017) 6\textendash 17

\bibitem{BrendanFilter} Brendan B. Godfrey, Jean-Luc Vay, "Improved numerical Cherenkov instability suppression in the generalized PSTD PIC algorithm", 
Computer Physics Communications, 196, 221 (2015) http://dx.doi.org/10.1016/j.cpc.2015.06.008


\bibitem{Lehe_Lagrangian}Remi Lehe, Manuel Kirchen, Brendan B. Godfrey,
Andreas R. Maier, and Jean-Luc Vay ``Elimination of numerical Cherenkov
instability in flowing-plasma particle-in-cell simulations by using
Galilean coordinates'' Phys. Rev. E 94, 053305 (2016).

\bibitem{Higuera} A. V. Higuera and  J. R. Cary, "Structure-preserving second-order integration of relativistic charged particle trajectories in electromagnetic fields", 
Physics of Plasmas 24, 052104 (2017)

\bibitem{Boris}J. P. Boris, \textquotedblleft Relativistic plasma
simulation-optimization of a hybrid code,\textquotedblright{} in Proceedings
of the Fourth Conference on Numerical Simulation Plas- mas Naval Research
Laboratory, Washington, D.C., 1970 , pp. 3\textendash 67.

\bibitem{VayPusher}J.-L. Vay ``Simulation of beams or plasmas crossing
at relativistic velocity'' Physics of Plasmas 15, 056701 (2008)

\bibitem{Esirkepov}T.Z.Esirkepov ``Exact charge conservation scheme for particle-in-cell simulations with an arbitrary form-factor'' Comp.
Phys. Communications \textbf{135}, p. 144-153 (2001).

\bibitem{Umeda2003}T.Y.Umeda, T. Omura, T. Tominaga, and H. Matsumoto "A new charge conservation method in electromagnetic particle-in-cell simulations'' Comp.
Phys. Communications \textbf{156}, p. 73-85 (2003).

\bibitem{Villacenor}John Villasenor and Oscar Buneman, ``Rigorous
charge conservation for local electromagnetic field solvers'' Comp.
Phys. Communications \textbf{69}, p. 306-316 (1992). https://doi.org/10.1016/0010-4655(92)90169-Y

\bibitem{Eastwood}James W. Eastwood, "The virtual particle electromagnetic particle-mesh method'' Comp.
Phys. Communications \textbf{64}, p. 252-266 (1991)

\bibitem{VLPL1999} A Pukhov "Three-dimensional electromagnetic relativistic particle-in-cell code VLPL (Virtual Laser Plasma Lab)"
Journal of Plasma Physics {\bf 61}, 425-433 (1999)

\bibitem{Bubble}A. Pukhov and J. Meyer-ter-Vehn, ``Laser wake field
acceleration: the highly non-linear broken-wave regime'' Appl. Phys.
B74 (2001) 355.

\bibitem{CowanEnvelope}Benjamin M. Cowan, David L. Bruhwiler, Estelle Cormier-Michel, Eric Esarey, Cameron G. R. Geddes, Peter Messmer, Kevin M. Pau "Characteristics of an envelope model for laser–plasma
accelerator simulation", Journal of Computational Physics {\bf 230}, 61-86 (2011)

\bibitem{Trojan} Hidding, B. et al. Ultracold electron bunch generation via plasma photocathode emission and acceleration in a beam-driven  plasma  blowout.  Phys.  Rev.  Lett.  {\bf 108},  035001 (2012).   

\end{thebibliography}


\end{document}